\documentclass[a4paper]{article}

\usepackage{INTERSPEECH2021}
\usepackage{graphicx,float}
\usepackage{lipsum}% Just for this example
\usepackage{graphicx}
\usepackage{enumitem}
\usepackage{subcaption}

\title{Fast Real-time Personalized Speech Enhancement:\\End-to-End Enhancement Network (E3Net) and Knowledge Distillation}
\name{Manthan Thakker$^*$\thanks{$^*$These authors contributed equally to this work.}, Sefik Emre Eskimez$^*$, Takuya Yoshioka, Huaming Wang}
\address{Microsoft Corporation, One Microsoft Way, WA, USA}
\email{\{mathakke, seeskime, tayoshio, huawang\}@microsoft.com}

\begin{document}

\maketitle
\begin{abstract} 

This paper investigates how to improve the runtime speed of personalized speech enhancement (PSE) networks while maintaining the model quality. Our approach includes two aspects: architecture and knowledge distillation (KD). We propose an end-to-end enhancement (E3Net) model architecture, which is $3\times$ faster than a baseline STFT-based model. Besides, we use KD techniques to develop compressed student models without significantly degrading quality. In addition, we investigate using noisy data without reference clean signals for training the student models, where we combine KD with multi-task learning (MTL) using an automatic speech recognition (ASR) loss. Our results show that E3Net provides better speech and transcription quality with a lower target speaker over-suppression (TSOS) rate than the baseline model. Furthermore, we show that the KD methods can yield student models that are  $2-4\times$ faster than the teacher and provides reasonable quality. Combining KD and MTL improves the ASR and TSOS metrics without degrading the speech quality. 

\end{abstract}
\noindent\textbf{Index Terms}: speech enhancement, noise suppression, personalized noise suppression, knowledge distillation, multi task learning, teacher student learning

\section{Introduction}  
\label{sec:intro}

Online conferencing tools have been widely adopted for conducting businesses and connecting with family and friends since the beginning of the COVID-19 pandemic. However, the background noise and reverberation degrade the speech and transcription quality. As these acoustic distortions can hamper communications and thus productivity, the speech enhancement (SE) field has drawn a lot of renewed attention recently~\cite{reddy2021interspeech,dubey2022icassp}. 

Personalized speech enhancement (PSE) provides an improvement to the general SE approach by using prior knowledge about a target speaker~\cite{dubey2022icassp,giri2021personalized,eskimez2021personalized,taherian2021one}. One exemplary approach to PSE is to extract a speaker embedding vector from a short enrollment audio sample of the target speaker and feed it to an SE model. 
This enables the SE model to remove interfering speakers in addition to the background noise. 
PSE provides practical benefits, especially for the users who share the same environments with other people during teleconferencing. 

To deploy the PSE models on various devices that are used for online meetings, it is paramount to make the computational cost very low while satisfying other performance requirements. The requirements include causal modeling for real-time processing, high-quality noise and interfering speaker suppression, little negative impact on the automatic speech recognition (ASR) accuracy, and little target speaker oversuppression (TSOS). Since the PSE models attempt to remove human voices spoken by the interfering speakers, they may occasionally suppress the target speaker too by mistake. Preventing TSOS is critical for applying the PSE models to real scenarios. 

In this paper, we examine multiple approaches to take on this challenge and build real-time PSE models with very low computational cost while maintaining satisfactory accuracy. We investigate this problem from three angles.
\begin{description}[style=unboxed,leftmargin=0cm]
    \item[End-to-end modeling:] We propose a personalized end-to-end enhancement network (E3Net), a  faster neural network model that is shown to improve the speech quality, WER, and reduce TSOS than a previously proposed  personalized deep complex convolution recurrent network (pDCCRN)~\cite{eskimez2021personalized}.
    \item[Knowledge Distillation (KD):] KD is one of the model compression approaches that are widely used in machine learning. We apply generic KD recipes for PSE by using a big capacity teacher model and a much smaller student model. We show the experimental results for both pDCCRN and E3Net. 
    \item[Leveraging unpaired noisy speech:] Most of the real-world recordings do not come with clean reference signals. Thus, PSE models are usually trained on clean-noisy speech sample pairs created by simulation, which may cause a mismatch between the training and deployment environments. We examine the effect of applying bigger teacher models to the real noisy data and using the outputs as clean references for student models. Combination with multi-task learning (MTL) using ASR transcriptions~\cite{eskimez2021personalized,eskimez2021human} is also considered. 
\end{description}

Our results show that E3Net provides better quality and runs $3\times$ faster than the pDCCRN model. Furthermore, our ablation study for E3Net suggests that using more filters on learnable encoder-decoder for the end-to-end network is required to obtain a high-quality model. In addition, the KD experiments show that student models as fast as $3\times$ their teacher models can provide reasonable quality for both pDCCRN and E3Net models. We show that we can further reduce TSOS with almost no  degradation to speech quality by combining MTL with KD methods.

\section{Related work} 

Personalized Speech Enhancement (PSE) is a speech enhancement method conditioned on a cue representing the target speaker. This cue is often provided as a static embedding vector that captures the target speaker's speech characteristics. The main goal of PSE is to filter out interfering speakers in addition to background noise and reverberation. 

Giri et al.~\cite{giri2021personalized} proposed Personalized PercepNet by modifying the original PercepNet model~\cite{valin2020percepnet} to accept speaker embeddings. Although the method was computationally efficient, it was based on a heuristic model and complex to build. The model was tested based on speech communication, and
the ASR accuracy was not considered. Eskimez et al.~\cite{eskimez2021personalized} proposed two PSE models, an evaluation metric called target speaker over-suppression (TSOS), and test sets to cover various scenarios. TSOS measures the degree of removal of the target speaker's speech segments and is critical for PSE since removing the target speech hampers effective conversations and degrades the transcription quality, as reported in~\cite{wang2020voicefilter}. Furthermore, Taherian et al.~\cite{taherian2021one} extended~\cite{eskimez2021personalized} to multi-channel scenarios by proposing a model that works with any microphone numbers and array geometries. Although the models of~\cite{eskimez2021personalized} can run on PCs in real-time, the computational cost was still too high for real usage as the audio processing can use only a tiny fraction of the available resources on devices. 

Knowledge Distillation (KD), or Teacher-Student (TS) learning, has been widely explored in natural language processing~\cite{sanh2019distilbert,turc2019well}, computer vision~\cite{wang2021knowledge,zhang2019your} and speech processing~\cite{chang2021distilhubert,peng2021shrinking} domains for training faster and more compact models with supervisions generated by computationally demanding teacher models. KD was first formalized in~\cite{bucilu2006model}. Hinton et al.~\cite{hinton2015distilling} trained a smaller student network with labels generated by a stronger teacher model, comprising an ensemble of models. Hao et al.~\cite{hao2020snr} used an SNR-based TS method to train an SE model using an ensemble of different models, which were individually trained for different SNR Ranges. Kobayashi et al.~\cite{kobayashi2021implementation} used KD to train a unidirectional recurrent student network with a bi-directional teacher network inspired by Born Again Networks~\cite{furlanello2018born}. Kim et al.~\cite{kim2021test} leveraged TS learning to perform online training of the student network for real-time speech enhancement on devices by using teacher's pseudo labels. The KD study in \cite{xu2022can} showed that KD helped avoid over-fitting with regularization, leading to more robust and generalized student networks. Recently, Chen et al.~\cite{chen2021ultra} explored various techniques, including objective shifting (OS) and layer-by-layer KD, for speech separation. 

In this paper, we first propose a real-time end-to-end PSE network with low computational cost, called E3Net.
The end-to-end modeling and optimization facilitate efficient use of the modeling capacity and easy development. 
We also apply the KD recipes to build more efficient PSE models and experimentally analyze their effectiveness.  In addition, we explore the use of KD to leverage unpaired noisy samples. Furthermore, we combine KD with multi-task training using speech recognition labels, which also allows the use of the unpaired noisy data~\cite{eskimez2021human,eskimez2021personalized}, to improve the transcription quality further. Extensive experimental results are shown to compare the various techniques. Note that model quantization or pruning, yet another popular approach to model compression, is out of the scope of our investigation as it relies heavily on run-time support. 

\begin{figure*}[t!]
  \centering
  \includegraphics[width=0.8\textwidth]{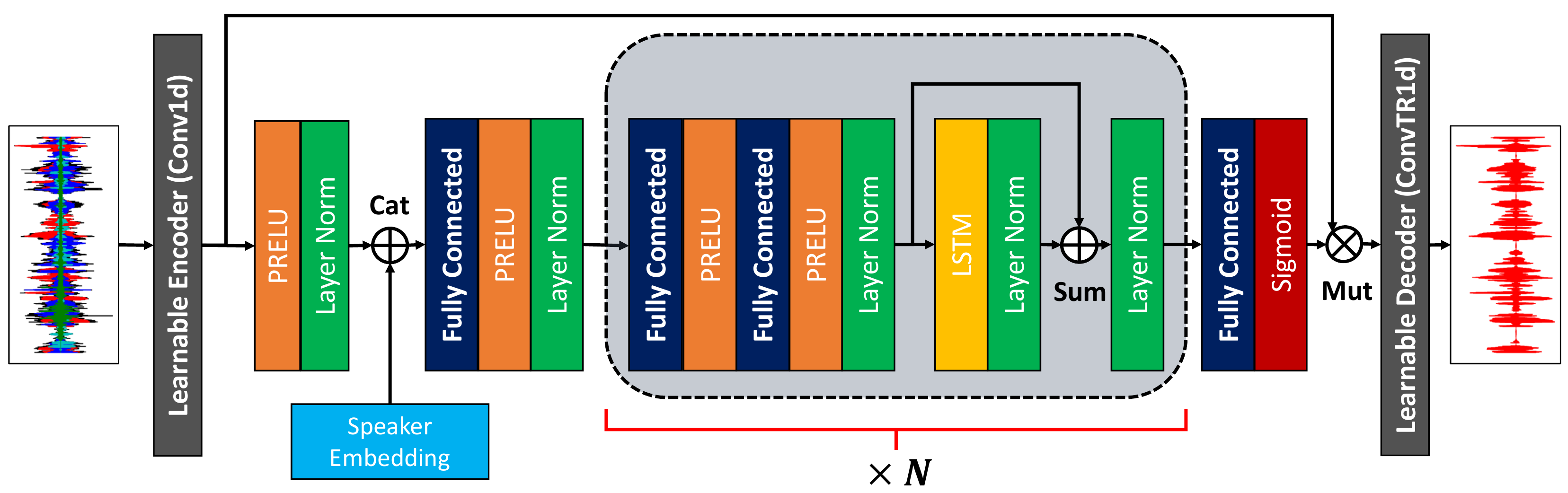}
  \vspace{-0.2cm}
  \caption{Proposed personalized E3Net model is shown. ``Cat'', ``Sum'' and ``Mut'' stands for concatenation, addition and element-wise multiplication. }
   \vspace{-0.5cm}
  \label{fig:E3Net}
\end{figure*}

\section{Proposed method} 

We approach the problem described in Section~\ref{sec:intro} from two different perspectives. First, we describe a simple and efficient end-to-end architecture that outperforms the previously proposed pDCCRN~\cite{eskimez2021personalized}. Next, we describe knowledge distillation and multi-task learning methods, which can be applied to any SE/PSE model architecture. 

\subsection{Personalized E3Net}  

We propose an end-to-end enhancement network (E3Net) that utilizes a learnable encoder and decoder instead of the short-time Fourier transform (STFT) features. 
Figure~\ref{fig:E3Net} shows the personalized E3Net model. 
The network takes a raw waveform as input and processes it using a 1D convolutional layer. Unlike some prior end-to-end enhancement models such as Conv-TasNet~\cite{luo2019conv}, we use filter and stride sizes that are equal to the window and hop sizes, respectively, of typical STFT configurations to reduce the computational cost. The learnable encoder extracts linear features and then feeds them to a PReLU non-linear activation layer and a layer normalization module. Subsequently, these features are concatenated with a speaker embedding vector and projected to an intermediate feature space by a linear layer plus non-linear activation. $N$ LSTM blocks process the intermediate features and feed them to a mask prediction layer which estimates masks to be applied to the original features with element-wise multiplication. The mask prediction layer consists of a fully connected layer with sigmoid non-linearity. Finally, a learnable decoder recovers a clean signal from the masked features. 

The LSTM blocks contain two fully connected layers with PReLU activation, where the second fully connected layer is followed by layer normalization. 
This two-layer fully connected block first maps the features to a higher dimensional non-linear space with the first layer and then converts them back to another space with the original dimensionality. 
Then, the output of this block is fed to a single LSTM layer, followed by another layer normalization module. We add the fully-connected block's output to this layer normalization output and apply another layer normalization module.

The end-to-end approach allows the whole network to be optimized without using redundant representations like the STFT features, facilitating the use of smaller models. Also, using only basic neural network elements such as fully connected and LSTM layers allows us to leverage optimized run-time modules for efficient computation. 

\subsection{Knowledge distillation} 
\label{sec:KD}

%The main goal of Knowledge distillation aka Teacher Student Learning is to train a faster, usually a smaller network with labels generated from a stronger network which is usually larger to distill the dark knowledge from the teacher pseudo Labels to the student which is not available by training with just ground truth labels. 

The main goal of KD or TS learning is to train a student model with labels or supervision signals generated by a teacher model. The student is typically smaller than the teacher. KD can also be used to leverage data without references, which are abundantly available. We investigate the following strategies for PSE:
% \begin{itemize}

\textit{Vanilla KD:} This method trains the student network with a supervised objective function where the teacher model output is used as the ground-truth label. These labels are called pseudo labels. The student model learns to mimic the teacher model. While we also examined Objective Shifting KD \cite{chen2021ultra} as well as starting KD with a pre-trained supervised model, their performance was similar to that of the vanilla KD. Therefore, we report only the vanilla KD results. % of which we present only results from Vanilla KD as results among these methods were similar.

%We also experimented with other KD recipes such as objective shifting, which %turned out to provide similar performance. Therefore, in the later experiment %section, we report only

\textit{Leveraging unlabeled data:} Real noisy speech data are usually obtained without the corresponding clean reference signals. To utilize these data, we explore using the unlabeled data with a strong teacher. Throughout the paper, we use the term ``unlabeled'' to refer to the data without a clean reference signal. Specifically, we construct two batches at each training step: 1) simulated noisy data with the corresponding ground-truth clean reference signal 2) real noisy data with pseudo labels generated by the teacher network.

\subsection{Multi-task learning with ASR loss function and KD}

Multi-task learning (MTL) using an ASR-based loss function is another approach for utilizing the unpaired real noisy signals for the training. 
One exemplary scheme is to alternate two model update steps. The first step, called SE-step, uses the conventional supervised loss function by using simulated data with the corresponding ground-truth clean signals. The second step, called ASR-step, computes an ASR-related loss by feeding the SE model output to a pre-trained sequence-to-sequence ASR model and updates only the PSE model parameters. This method was shown to work with different ASR models for non-personalized~\cite{eskimez2021human} and personalized SE~\cite{eskimez2021personalized}.

%Our previous works~\cite{eskimez2021human,eskimez2021personalized} show the benefit of multi-task learning (MTL) with a frozen ASR back-end and only updating the front-end model. We consider two alternating steps to calculate different loss values 1) SE-step: we calculate the supervised loss for PSE and update the PSE network, 2) ASR-step: we compute the ASR loss and update the PSE network. In practice, this method makes the front-end models more friendly to any ASR engine, as we showed results on a different ASR engine. 

% We consider applying the MTL to KD's student models with two different approaches. In the first approach, we apply the same technique described in our previous work. We train the student network with alternating steps for ASR and PSE loss. In the second approach, we add a third step: we extract the pseudo labels from the unlabeled data and apply supervised loss as we described in Section~\ref{sec:KD}.
We consider applying the MTL to KD's student models using the same technique described in our previous work. We train the student network with alternating steps for ASR and PSE loss. Different from the previous version, we add a third step: extracting the pseudo labels from the unlabeled data using the teacher model and applying supervised loss using these pseudo labels for the student model.

\begin{table*}[ht!]
  
  \caption{Experimental results for the VCTK dataset using three testing scenarios. TS1 includes target, interfering speaker, and noise, TS2 includes target speaker and noise, and TS3 includes only the target speaker. All conditions include reverberation. For WER and TSOS metrics, the lower is better. For the DNSMOS metric, the higher is better. $M$ stands for millions. TSOS is described in seconds. RTF is computed using a single-thread (no parallelization) Intel(R) Xeon(R) W-2133 CPU @ 3.60GHz, averaged over 100 runs. $\mathcal{L}_{KDonSup}$ and $\mathcal{L}_{KDonUnlab}$ models are trained with pseudo labels obtained from simulation and unlabeled data, respectively.}
  \vspace{-2mm}
  \centering
 \resizebox{\textwidth}{!}{
\begin{tabular}{llrcccccccccccc}                                                                                           \cline{2-15} 
\multicolumn{1}{c}{} && \multicolumn{2}{c}{Complexity} && \multicolumn{3}{c}{TS1} && \multicolumn{3}{c}{TS2} && \multicolumn{2}{c}{TS3}                      \\ \cline{2-4} \cline{6-8} \cline{10-12} \cline{14-15} 
\multicolumn{1}{c}{}                       &  & \multicolumn{1}{c}{Parameters} & \multicolumn{1}{c}{RTF} && \multicolumn{1}{c}{WER}  & \multicolumn{1}{c}{DNSMOS} & \multicolumn{1}{c}{TSOS} &  & \multicolumn{1}{c}{WER} &  \multicolumn{1}{c}{DNSMOS}  & \multicolumn{1}{c}{TSOS} &  & \multicolumn{1}{c}{WER} & \multicolumn{1}{c}{TSOS} \\ \cline{1-15}
No Enhancement              && - & - && 43.03 & 2.92 & 0.00 && 13.35 &	2.98  & 0.00 && 7.12 & 0.00 \\ \hline
pDCCRN-baseline (Teacher)                       && 3.96 M & 0.195 && 38.24 & 3.36 & 14.61 && 22.07 & 3.61  & 10.46 && 13.10 & 4.20 \\ \hline
pDCCRN-student  \\                    
\hspace{0.4cm} - $\mathcal{L}_{supervisedSE}$  && 1.50 M & 0.094 && 44.36 & 3.22  & 51.89 && 24.43 & 3.52 & 21.75 && 13.20 & 19.50 \\ 
\hspace{0.4cm} - $\mathcal{L}_{KDonSup}$  && 1.50 M & 0.094 && 38.93 & 3.37 & 4.83 && 21.31 & 3.57 & 2.70 && 10.57 & 3.98 \\
% \hspace{0.4cm} - $\mathcal{L}_{objectiveKD}$  && 1.50 M & 0.03 & 39.84 & 10.05 & 3.29 & 9.45 && 22.62 & 6.36 & 3.52  & 7.97 && 2.87 & 7.94 \\
% \hspace{0.4cm} - $\mathcal{L}_{SupervisiedInit+KD}$  && 1.50 M & 0.03 &  38.41  & 8.61 & 3.38	 & 13.61 && 21.44 & 5.58 & 3.60  & 9.98 && 3.24 & 7.94 \\ 
% \hline
% pDCCRN student MultiTask - Training                  \\  
\hspace{0.4cm} - $\mathcal{L}_{KDonUnlab}$  && 1.50 M & 0.094 && 38.31 & 3.36 & 10.66 && 21.93 & 3.57 & 9.14 && 9.31 & 4.76  \\ 
% \hspace{0.4cm} - $\mathcal{L}_{vanillaMTL}$  && 1.50 M & 0.03 & 37.03 & 5.48 & 3.32 & 0.45 && 17.41 & 2.95 & 3.49 & 0.28 && 0.83 & 0.56  \\
\hspace{0.4cm} - $\mathcal{L}_{MTL+KDonUnlab}$  && 1.50 M & 0.094 &&  37.20 & 3.37  & 0.60 && 17.81 & 3.55 & 0.58 && 7.82 & 0.31 \\
\hline
E3Net-baseline ($N=4$)                 && 6.61 M & 0.065 && 37.94 & 3.46 & 3.75 && 19.68 & 3.68 & 1.83 && 8.13 & 3.32\\ \hline
% E3Net (4 Layers) @ 1.3M                && 6.61 M & 0.04 & 32.53 & 4.36 & 3.58 & 1.48 && 16.42 & 2.34 & 3.80 & 0.49 && 0.95 & 3.97 \\ \hline
E3Net (Teacher) ($N=8$)             
 && 10.85 M & 0.122 &&  38.19 & 3.48 & 2.60 &&
18.68 & 3.71 &	1.01 &&
7.69& 0.88\\ \hline
E3Net-student ($N=2$)                      \\
\hspace{0.4cm} - $\mathcal{L}_{supervisedSE}$ && 4.50 M & 0.036  && 42.52 & 3.38 & 3.57 && 21.55 & 3.61 & 1.68 && 8.76 & 1.73 \\ 
\hspace{0.4cm} - $\mathcal{L}_{KDonSup}$  && 4.50 M & 0.036 && 41.89 & 3.42 & 2.99 && 19.79 & 3.65 & 0.56 && 7.73 & 0.18 \\
% \hspace{0.4cm} - $\mathcal{L}_{objectiveKD}$  && 4.50 M & 0.018 &  42.09 & 5.93 & 3.41 & 2.92 && 19.92 & 2.93 & 3.63 & 0.67 && 2.07 & 4.02 \\
% \hspace{0.4cm} - $\mathcal{L}_{SupervisiedInit+VanillaKD}$  && 4.50 M & 0.018 &  40.94 & 7.55 & 3.41 & 8.26 && 20.25 & 4.16 & 3.67 & 4.24 && 2.54 & 4.00  \\ \hline
% E3Net student - MultiTask Training \\
% \hspace{0.4cm} - $\mathcal{L}_{vanillaMTL}$ && 4.50 M & 0.018  & 42.23 & 4.42 & 3.45 & 0.80 && 19.65 & 2.80 & 3.63 & 0.43 && 0.79 & 3.89 \\ 
\hspace{0.4cm} - $\mathcal{L}_{KDonUnlab}$  && 4.50 M & 0.036 && 41.83 & 3.48 & 0.56 && 19.90 & 3.65 & 0.28 && 7.50 & 0.16  \\ 
\hspace{0.4cm} - $\mathcal{L}_{MTL+KDonUnlab}$ && 4.50 M & 0.036  && 42.50 & 3.47 & 0.50 && 19.48 & 3.64 & 0.27 && 7.51 & 0.13 \\ 
 \hline
 \end{tabular}
}
\label{tab:main_res}
\vspace{-0.5cm}
\end{table*}

\begin{figure}[t!]
\centering
  \centering
	\begin{subfigure}[b]{0.9\columnwidth}
\includegraphics[width=\columnwidth]{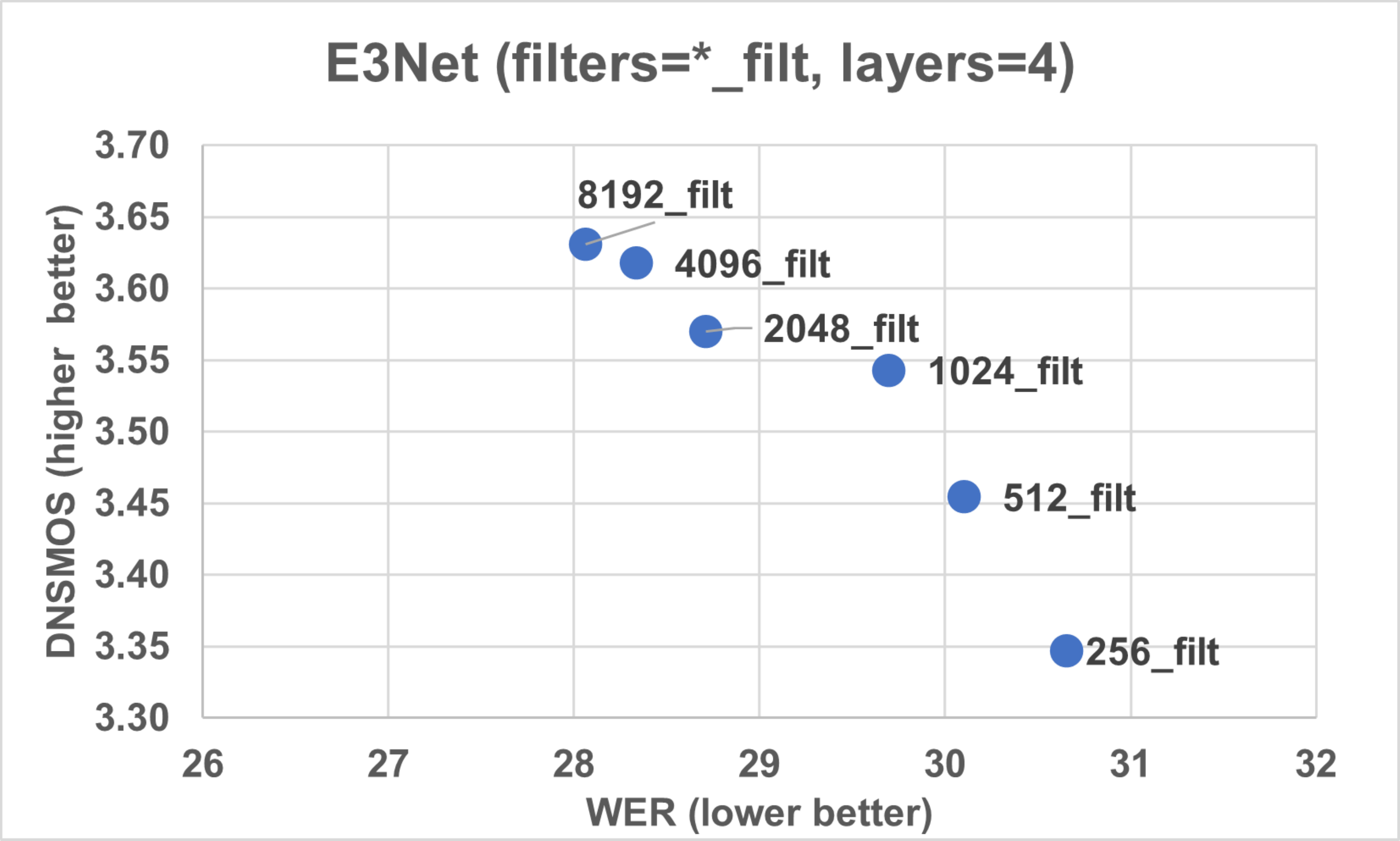}
 		\vspace{-1em}
\caption{}\label{fig:E3Net_abl_a}
	\end{subfigure}
	\hfill
	\begin{subfigure}[b]{0.9\columnwidth}
\includegraphics[width=\columnwidth]{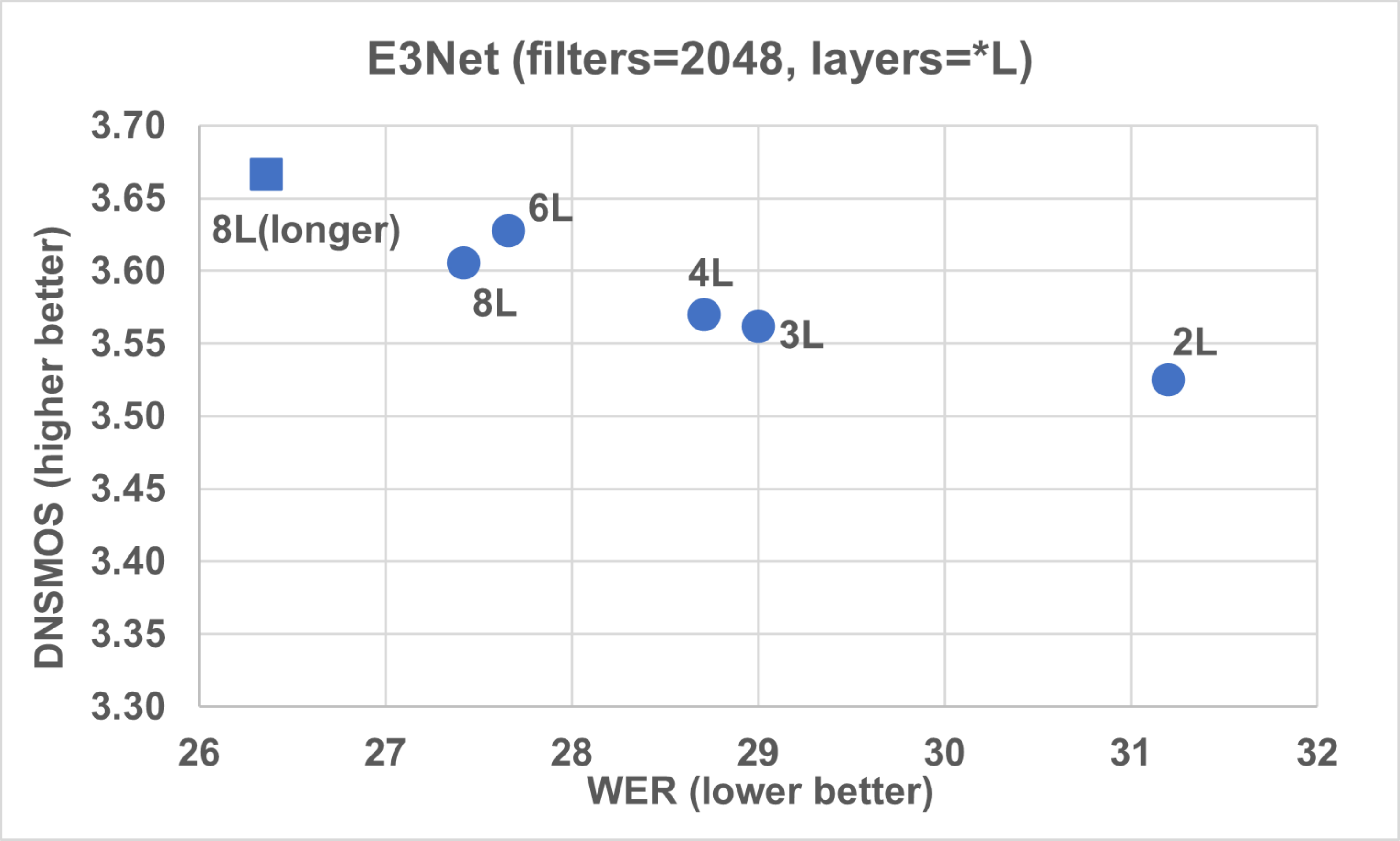}
 		\vspace{-1em}
\caption{}\label{fig:E3Net_abl_b}
	\end{subfigure}
	\vspace{-1em}
  \caption{E3Net performance is shown for different numbers of encoder/decoder filters and LSTM block numbers $N$}
  \label{fig:E3Net_abl}
  \vspace{-2em}
\end{figure}

\section{Experiments}
% This section reports our experimental setups and results. %First, we describe our training and evaluation data. Then, we show the implementation details and evaluation metrics. Finally, we analyze the results.

\subsection{Experimental settings}
We followed the same data configurations as those employed in~\cite{eskimez2021personalized}, using 2000 hours of training and 50 hours of validation data. 
The clean speech signals for simulation were taken from the DNS Challenge~\cite{reddy2021interspeech}, comprising high-quality LibriVox speech data that were selected based on mean opinion score (MOS) values. The noise files were obtained from AudioSet~\cite{gemmeke2017audio} and Freesound~\cite{fonseca2017freesound}. Simulated room impulse responses (RIRs) were used to control the target and interfering speakers' distances to the microphone. 
In addition, we used a subset (1302 hours) of 75K hours ASR training set described in~\cite{eskimez2021human} for the experiments requiring unpaired noisy data as it contained ASR reference labels and thus was usable for both the multi-task learning and KD experiments. 

For evaluation, we used our previous test set~\cite{eskimez2021personalized} that included three scenarios: 1) \textbf{TS1}: target speaker + interfering speaker + noise, 2) \textbf{TS2}: target speaker + noise, and 3) \textbf{TS3}: target speaker only. We constructed these sets based on the VCTK corpus~\cite{veaux2017superseded}. We simulated noisy speech samples for each file, then concatenated all the files of the same target speaker to obtain a single long-duration test file for each speaker. This test set was challenging as the noise and reverberation characteristics changed every few seconds. The average duration of the files was 27.5 minutes.

%\subsection{Implementation Details}
In addition to the proposed E3Net, we also used pDCCRN to obtain widely applicable insights. For pDCCRN, we used the original setting of \cite{eskimez2021personalized} for the teacher model: we set the numbers of filters in the encoder and decoder 2D convolutional layers to [16, 32, 64, 128, 128, 128], the kernel sizes to $5\times2$ and the strides to $2\times1$. For the student model, we set the filter numbers to [8, 16, 32, 32, 64, 64] and retained the kernel sizes and strides. For both configurations, the LSTM hidden size was 128. We used the cosine annealing learning rate scheduler with a peak learning rate of $10^{-3}$. The batch size for the pDCCRN models was 8 with a gradient accumulation of 2. For STFT, we used a 32 ms window size and a 16 ms hop size.

For E3Net, we set the number of LSTM blocks to 4, the number of features for the learnable encoder and decoder to 2048, the embedding and LSTM dimensions to 256, and the hidden dimension of the fully connected block to 1024. We used this baseline configuration for the teacher model, except the number of LSTM blocks was 8 instead of 4. For the student, we set the number of LSTM blocks to 2. In addition, we conducted parameter sweeping experiments to show the impact that the model configurations have on the E3Net performance. Gradient accumulation was not used for the supervised and KD training since the memory footprint was much smaller with E3Net. We used the same learning rate scheduler as pDCCRN except for the peak learning rate of $10^{-4}$ instead of $10^{-3}$. The window and hop sizes were set to 20 ms and hop size to 10 ms, respectively, for E3Net, which helps further reduce the processing latency. 

For MTL experiments, we loaded supervised models trained for 209K updates and resumed the training with a peak learning rate of $5\times10^{-5}$. All other configurations were the same as the supervised training. For KD experiments, we used similar settings, except for choosing the best checkpoint for each model, which was typically between 200-300K updates for pDCCRN and between 300-500K updates for E3Net. We used gradient accumulation of 2 for all the KD experiments with the same batch size as the supervised experiments.

The performance of the models was assessed with the metrics described in~\cite{eskimez2021personalized}. Specifically, we measured the speech and transcription quality, target speaker over-suppression (TSOS), and real-time factor of the models. For speech quality, we used DNSMOS P.835~\cite{reddy2021dnsmos}, which is a neural-network-based mean opinion score (MOS) estimator that shows a good correlation with subjective ratings. We used the word error rate (WER) to measure the transcription quality. TSOS estimates the removed target speaker segments, and thus it is critical for the models to have very low TSOS values to avoid creating disruption during the usage in online meetings. Unlike our previous work, TSOS was normalized to show the target-speaker-suppressed period in seconds per half an hour. 

\subsection{Results}
Table~\ref{tab:main_res} shows the main results, including those of KD and MTL. First, let us focus on the comparison between E3Net and pDCCRN. Our experiments find E3Net with $N=4$ had a fair balance between RTF and quality. Therefore, we chose this configuration as our baseline. In the table, compared to the pDCCRN-baseline, not only did E3Net-baseline provide better results for all the metrics and test scenarios, but it also ran $3\times$ faster (0.195 to 0.065). Notably, the TSOS improvement was substantial for TS1 (14.61 to 3.75 seconds) and TS2 (10.46 to 1.83 seconds) scenarios. 

Figure~\ref{fig:E3Net_abl} shows the results of the parameter sweeping experiments for E3Net. 
These plots show the average DNSMOS and WER values computed over all three test sets. We investigated the impacts of two hyperparameters while keeping  the other parameters fixed: (1) the number of filters in the learnable encoder-decoder; and (2) the number of LSTM blocks ($N$). Figure~\ref{fig:E3Net_abl_a} shows that using more filters in the learnable encoder-decoder improved the results significantly. With a small number of filters (256 or 512) as with the STFT feature dimension, E3Net yielded poor results. Figure~\ref{fig:E3Net_abl_b} shows that using more LSTM blocks also improved the results, especially WER. In addition, as implied by the E3Net result with $N=8$ trained for additional 100K iterations (shown by a square in the plot), we noticed that training E3Net significantly longer could yield substantial gains irrespective of the model configurations, which was not observed for the STFT-based models like pDCCRN. While we limit the maximum iteration number for a fair comparison with pDCCRN, this E3Net behavior deserves future investigation. From these results, we can conclude that both over-parameterization for the encoder-decoder and using more LSTM blocks significantly improve the performance of E3Net. 
%Also, it seems E3Net can benefit from training much longer than the STFT-based models. 

%We performed Objective Shifting KD, Vanilla KD and resuming KD from Supervised %model, of which we presented results from the best recipe in %Table~\ref{tab:main_res} as results were similar.
%For KD Experiments, 
Table~\ref{tab:main_res} shows that the smaller student models for both architectures resulted in significant degradation compared to the baseline models when trained with only the supervised loss ($\mathcal{L}_{LKDonSup}$). Models trained using KD provided better TSOS values while improving WER and DNSMOS. 
% However, all student models trained with KD outperformed the student trained with supervised loss and were much closer to the teachers, if not better in terms of quality for some metrics. %We tried different KD recipes namely, Objective Shifting \cite{chen2021ultra} and resuming KD on pre-trained SE model, the results are similar to KD and we only present the best results.
The networks trained with pseudo labels on the unlabeled data ($\mathcal{L}_{KDonUnlab}$) provide similar improvements to KD ($\mathcal{L}_{LKDonSup}$), which indicates that 
the performance gains can be largely attributed to mimicking the behavior of the teacher models rather than the KD data. However, it is noteworthy that the E3Net-student with $\mathcal{L}_{KDonUnlab}$ model provides significant improvements to TSOS compared to its pDCCRN counterpart. 

%Furthermore, vanilla MTL improves the TSOS significantly for both models. It also improves WER for both models, especially for TS2. Adding the KD loss into MTL, denoted as $MTL+KD$ in the table, does not increase performance compared to vanilla MTL for both models and all three scenarios. $MTL+KD$ might have added robustness to the models, and to measure this, we need to evaluate the models on the non-simulated data, which is our future work. 

Combining MTL and KD improved the TSOS and WER significantly for both models ($\mathcal{L}_{MTL+KDonUnlab}$). This was achieved without causing degradation to DNSMOS. In our prior work~\cite{eskimez2021personalized}, it was observed that MTL slightly degraded the DNSMOS. Combining KD and MTL might have helped regularize the training to prevent the degradation, although further investigation is desired to draw firm conclusions.

\section{Conclusions}
In this work, we proposed a novel fast end-to-end personalized speech enhancement (PSE) architecture named E3Net. In addition, we applied knowledge distillation (KD) to E3Net and STFT-based pDCCRN models to compress them and reduce their computational cost while maintaining reasonable quality. Furthermore, we used large-scale unlabeled data to train a student network using larger teacher models' pseudo labels. Our results showed that E3Net outperforms pDCCRN with a $3\times$ faster real-time factor (RTF). In addition, results showed that KD recipes could compress the models further ($2-4\times$) with a slight degradation to the quality.  

\newpage

\bibliographystyle{IEEEtran}

\bibliography{main}

\end{document}